\documentclass{natureprintstyle}
\usepackage{color}
\usepackage[normal,normal,bf]{caption}

\usepackage{amsfonts}
\usepackage{amssymb}
\usepackage{amsmath}

\usepackage{float}
\usepackage{placeins}
\usepackage{graphicx,comment}
\usepackage[export]{adjustbox}
\makeatletter
\def\fnum@figure{{\bf Fig.~\thefigure}}
\makeatother

\makeatletter
\let\saved@includegraphics\includegraphics
\AtBeginDocument{\let\includegraphics\saved@includegraphics}
\makeatother

\usepackage[export]{adjustbox}

\usepackage[super,comma,sort&compress]{natbib}

\usepackage{hyperref}
\usepackage{verbatim}

\newcommand\araa{Ann. Rev. Astron. Astrophys.}
\newcommand\apj{Astrophys. J.} 
\newcommand\apjl{Astrophys. J. Lett.} 
\newcommand\mnras{Mon. Not. R. Astron. Soc.} 
\newcommand\aap{Astron. \& Astrophys.} 
\newcommand\pasj{Publ. Astron. Soc. Jpn}
\newcommand\pasp{Publ. Astron. Soc. Pac}

\newcommand{\chandra}{{\emph{Chandra}}}

\newcommand{\nustar}{{NuSTAR}}
\newcommand{\xrism}{{XRISM}}

\newcommand{\gx}{GX~13+1}
\newcommand{\rxte}{\textit{RXTE}}

\title{Stratified wind from a super-Eddington X-ray binary is slower than expected}

\author{XRISM collaboration$^*$}

\begin{document}

\maketitle
\footnotetext[1]{
A list of participants and their affiliations appears at the end of the paper.
}

\begin{abstract}

Accretion discs in strong 
gravity ubiquitously produce winds, 
seen as blueshifted absorption lines in the X-ray band of both stellar mass X-ray binaries 
(black holes and neutron 
stars)\cite{Ponti2012, 
DiazTrigo2016,Neilsen2023,Parra2024} 
and supermassive black 
holes\cite{Nardini2015}. Some of the 
most powerful winds (termed Eddington winds)
are expected to 
arise from systems where radiation 
pressure is sufficient to unbind 
material from the inner disc 
($L\gtrsim L_{\rm Edd}$). These winds should be extremely 
fast and carry a large amount of 
kinetic power,  which, when associated with supermassive black holes, would make them a prime contender for the feedback mechanism linking the growth of those black holes with their host galaxies
\cite{KP2015}. 
Here we show the XRISM Resolve 
spectrum of the Galactic 
neutron star X-ray binary, GX 13+1, which reveals one of the densest
winds ever seen in absorption lines. This Compton-thick wind significantly attenuates the flux, making it appear faint, although it is intrinsically
more luminous than usual ($L\gtrsim L_{\rm Edd}$). 
However, the wind is extremely slow, more
consistent with the predictions 
of thermal-radiative winds launched 
by X-ray irradiation of the outer 
disc than with the expected 
Eddington wind driven by 
radiation pressure from the inner disc. 
This puts new constraints on the 
origin of winds from bright 
accretion flows in binaries, but 
also highlights the very different 
origin required for the ultrafast ($v\sim 0.3c$) winds seen in recent Resolve observations of a supermassive black hole
at similarly high Eddington 
ratio\cite{xrism2025}.

\end{abstract}


\gx\ is a disk-accreting 
neutron star in a 24.5-day orbit\cite{Corbet2010,Iaria2014} with a giant (K5 III) companion star\cite{Bandyopadhyay1999}, which gives a large mass transfer rate through the Roche lobe\cite{Coriat2012}, resulting in a persistently bright X-ray source ($L\approx 0.5L_{\rm Edd}$ for a $1.4M_\odot$ neutron star at a distance of 7~kpc; ref. \citenum{Bandyopadhyay1999}).
The presence of dips in the X-ray lightcurve indicates a high binary inclination\cite{D'Ai2014}, which is optimal for observations of accretion disc winds\cite{Boirin2003,DiazTrigo2006,Ponti2012}.
Every X-ray observation of \gx\ with sufficient spectral resolution has shown blueshifted absorption lines\cite{Ueda2004, DiazTrigo2012, Madej2014, Allen2018}, making it an ideal target for Resolve, an X-ray micro-calorimeter onboard the new JAXA/NASA/ESA mission \xrism\cite{Tashiro2025}. Resolve has an energy resolution of 4.5 eV at 6~keV, which is a factor 4 better than the previous state of the art equipment for bright binaries (third order data\cite{Miller2016} from the High-Energy Transmission Grating Spectrometer\cite{Canizares2005}, hereafter HETGS), and with much larger effective area, especially above 7~keV. The combination of improved resolution and larger area enables more sensitive measurements of the velocity and ionisation structure of accretion disc winds, critical for diagnosing the physical properties and launch mechanisms of these outflows\cite{Neilsen2013}. 

\begin{figure*}[t]
\centering
\includegraphics[width=0.8\hsize]{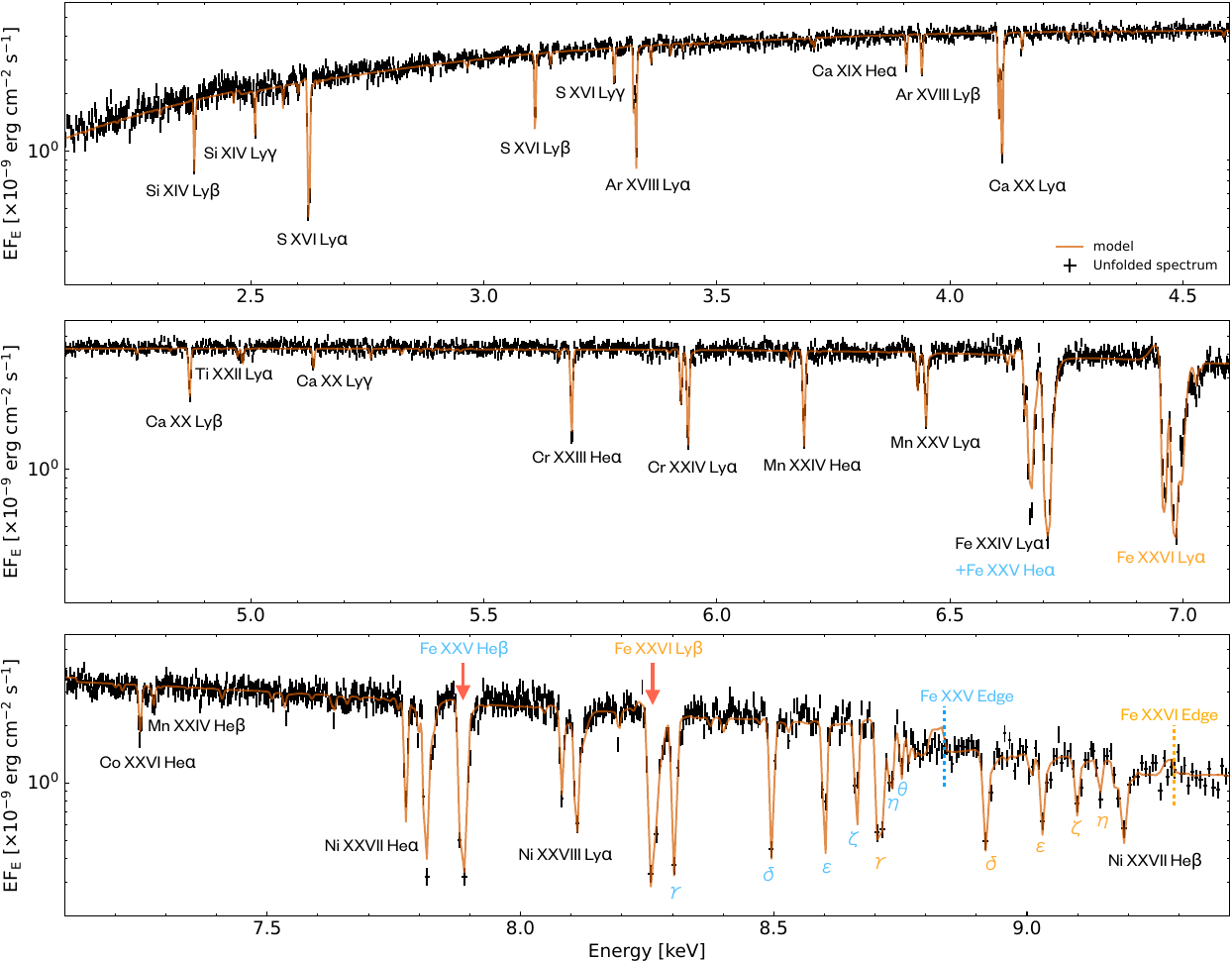}
\begin{spacing}{1.0}
\caption{\textbf{\label{fig:resolve-label} The Resolve/{\it XRISM} spectrum of \gx.} This is dominated by multiple absorption lines from H- and He-like ions, blueshifted by $\sim 330$~km/s. All the fine structure transitions in these lines are resolved, showing that the lines are very narrow (velocity dispersion $\sim 150$~km/s).
All strong lines are labelled; the $1-n$ transitions of Fe\,{\sc xxv} and Fe\,{\sc xxvi} are indicated in cyan and orange, respectively. Even the weakest line identified here (Ti XXII Ly$\alpha_{1,2}$ around 4.95~keV) is highly significantly detected ($\Delta\chi^2=32$ for 1 additional degree of freedom). The orange line shows the best fit model described in the text, with an intrinsic continuum absorbed by the slow wind, 
but with a faster (700km/s), broader (300km/s) even more highly ionised component to fit the blue wing seen in Fe\, {XXVI} Ly$\alpha_{1,2}$
(see Figure~\ref{fig:resolve-zoom}). The model also includes diffuse emission from the wind (modelled using scattered intrinsic flux plus photoionised line and recombination continuua from both wind components, all with some self absorption in the wind). This fits the data fairly well overall (see Methods, Extended Data Table \ref{tab:2pion_2emit_XR}), except around the FeXXV (8.8~keV) and FeXXVI (9.25~keV) edges where the photoionisation model used here is incomplete (only including transitions up to $n=16$). The total electron scattering optical depth in the slow(both) winds is $\tau_{\rm es}\sim 1(1.8)$, attenuating the intrinsic flux (see Figure~\ref{fig:archive}).}
\end{spacing}
\end{figure*}

The new data on \gx\ from Resolve were 
taken on 25 February 2024; more 
details of the observations and data 
analysis are given in Methods. The 
Resolve spectrum, shown in 
Figure~\ref{fig:resolve-label}, reveals dozens of strong, slightly blueshifted 
($v_{\rm out}\approx 330$~km/s), narrow 
($v_{\rm turb}\approx 150$~km/s) X-ray 
absorption lines. Most of these are 
from H-like and He-like ions of 
multiple elements (Si, S, Ar, Ca, Ti, 
Cr, Mn, Fe, Co, Ni), indicating a 
highly ionised absorber. 

Many of the lines below 7~keV (Figure~\ref{fig:resolve-label}a,b) have been studied before, though at lower resolution and signal-to-noise, in \chandra\ HETGS observations of this source\cite{Allen2018}. What is completely new is the number and depth of lines above 7~keV (Figure~\ref{fig:resolve-label}c), with multiple higher order transitions detected out to at least K$\theta$ ($n=1-9$) for Fe\,{\sc xxv} and K$\eta$ ($n=1-8$) for Fe\,{\sc xxvi}. These lines have small oscillator strengths, so their depth 
requires the column density of the wind to be extremely high, both in an absolute sense and in comparison to the prior \chandra\ observations (see below). 

We first model the absorption lines from each ion separately, and find that these are consistent with being 
produced by a single absorber with ionisation parameter $\log \xi\approx 3.9$ and an equivalent column density of $N_{\rm H}\approx 1.3-1.4\times 10^{24}$~cm$^{-2}$, assuming solar abundances. We confirm this by modelling all the ions together using the photoionisation code {\sc pion} (refs. \citenum{Kaastra1996,Mehdipour2016}) (see Methods). The derived 
column is so large that the wind is optically thick to electron scattering, with $\tau_{es}\approx 1$. This attenuates the radiation from the central X-ray source as it passes through the wind, reducing the direct continuum flux by a geometry-dependent factor $\sim \exp(\tau_{es})\approx 3$.
Correcting for this effect, we infer a bolometric luminosity of $L=1.8\times10^{38}$ erg s$^{-1}$, which is approximately $ L_{\rm Edd}$: this Compton-thick wind is produced by a source radiating at the Eddington limit. 

Both these conclusions were 
initially surprising, as the ionised winds 
seen in previous observations of \gx\ with the 
\chandra\ HETGS had large but optically thin 
column densities\cite{Ueda2004, Allen2018} of 
$N_{\rm H}\approx 2-3\times 10^{23}$~cm$^{-2}$, 
from a source with large but sub-Eddington 
luminosities of about $ 0.5L_{\rm Edd}$. The 
unusual behavior of GX13+1 during our 
observation is very clearly illustrated by our 
simultaneous \nustar\ broadband X-ray data (Methods). 
These are shown as the orange points 
in Figure~\ref{fig:archive}, in which we also show an 
archival \nustar\ spectrum (green). 
GX13+1 is noticeably fainter during our 
observation and has a much stronger absorption 
feature at $\sim 8.8$ keV, that is, at the K-edge 
of Fe\, {\sc xxv}. 
The column 
density in this ion is plainly much larger 
than in the majority of archival data. Other 
archival broadband datasets\cite{Homan2004} 
(RXTE: grey) confirm that this ``reduced 
flux/strong Fe\, {\sc xxv} edge" state is very 
unusual ($\lesssim10\%$ of observations), but 
lack the energy resolution to unambiguously connect this 
behavior to attenuation in a high-column wind.

\begin{figure}[t]
\includegraphics[width=\linewidth]{archive.eps}\\
\vspace{-5mm}
\begin{spacing}{1.0}
\caption{\textbf{\label{fig:archive} Historical X-ray variability of \gx.} The archival \nustar\ spectrum of \gx\ (green) is similar to most of the archival \rxte\ data (grey). Instead, the \xrism-coordinated \nustar\ spectrum (orange) has lower flux and shows a much deeper K edge from Fe\,{\sc XXV} at 8.8~keV. On closer inspection, 5-10\% of the archival \rxte\ spectra are similar to this recent \nustar\ observation, indicating that this
dense wind/super-Eddington phase 
is recurring in the source. The blue band shows a range of possible continuum spectra of \gx\ after correcting for attenuation due to electron scattering in the wind. The lower end of the envelope corresponds to $\tau_{\rm es}=1$ from the slow wind alone, while the upper end corresponds to 
$\tau_{\rm es}=1.8$ as inferred from the best fit model for the slow plus fast wind. The source is intrinsically more luminous than normal, at or above Eddington.}
\end{spacing}
\end{figure}

\begin{figure}[t]
\centering 
\includegraphics[width=\linewidth]{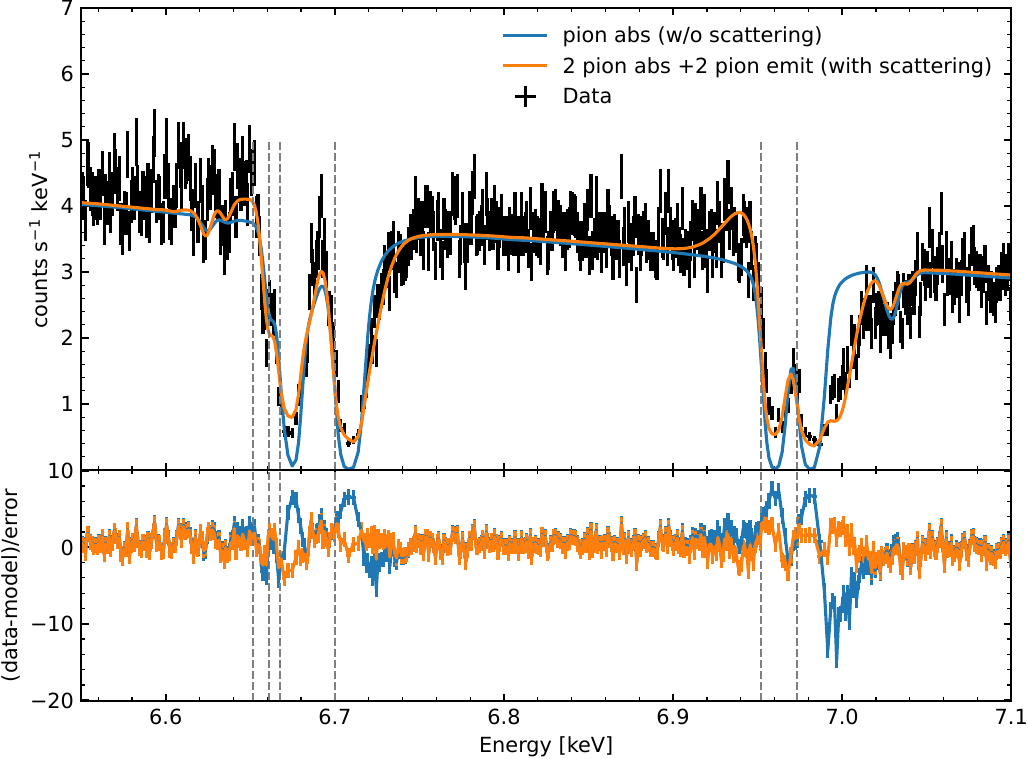}
\vspace{-5mm}
\begin{spacing}{1.0}
\caption{\textbf{\label{fig:resolve-zoom} Zoom in around the Fe K$\alpha$ lines}. The
dashed vertical lines show the restframe energies of (from left to right) the weak 
Fe {\sc xxiv} doublet (6.652 and 6.661 keV), Fe {\sc xxv} (intercombination: 6.667 
 and resonance: 6.700 keV)
and the Fe {\sc xxvi} doublet (6.952 and 6.973 keV). The blue line shows a single 
photoionised absorption model with parameters which fit the multiple narrow lines in the rest of the spectrum. This predicts that the lines go black in their centres, but the data show residual emission due to the presence of diffuse flux (most likely reprocessing and scattering from the wind itself). It also misses the blue wing in the Fe {\sc xxvi} Ly$\alpha1,2$ absorption line at 7~keV, showing that there is higher velocity material at higher ionisation state. The orange line shows our best model including these addtional components in both absorption and emission.  
}
\end{spacing}
\end{figure}

Correcting our best-fit continuum model for electron scattering attenuation in the slow wind gives an intrinsic flux shown as the lower edge of the blue band in Figure~\ref{fig:archive}, implying the source is intrinsically more luminous than normal. We suggest a causal relationship: an increase in the intrinsic X-ray luminosity enhances the wind to such an extent that it becomes Compton-thick, suppressing the observed flux and making the source appear dimmer (but with strong wind signatures).
The increase in wind column may also explain the different X-ray polarisation properties seen by IXPE during the \xrism\ observation\cite{Bobrikova2024}.

The optically thick column in the slow wind in \gx\ is actually only part of the material obscuring the source. Figure~\ref{fig:resolve-zoom} shows a detailed view of the strongest
lines, the  Fe~{\sc XXV} K$\alpha$ intercombination ($x+y: 6.670$~keV) and resonance ($w: 6.700$~keV) transitions, and the Fe~{\sc XXVI} Ly$\alpha_1,\alpha_2$ fine structure doublet. The blue line shows the predicted absorption line profiles using the column and velocities derived above for the slow wind, assuming that this covers all of the intrinsic X-ray emission region(s). 
It is immediately apparent that the model predicts that 
the narrow cores of these strong lines should be completely black (opaque) at their centres, whereas the data 
unambiguously show residual flux. 
This shows there must be an additional source of X-rays that is not absorbed by the wind, most likely the wind itself, which scatters and re-radiates some of the X-rays from the central source, forming a  diffuse secondary source of X-rays.
The comparison 
also shows that the 
single velocity absorption model misses the blue (higher-energy) wing seen especially on the
Fe~{\sc XXVI} Ly$\alpha_{1}$ absorption line at around 7~keV, 
indicating the presence of higher velocity material. 
This most likely indicates that the wind is stratified, with the most highly ionised material having speeds that are roughly twice that of the slower, less ionised material that forms the narrow line core.

\begin{figure}[t]
\vspace{-8mm}
\hspace{-8mm}
\includegraphics[width=1.1\hsize,left]{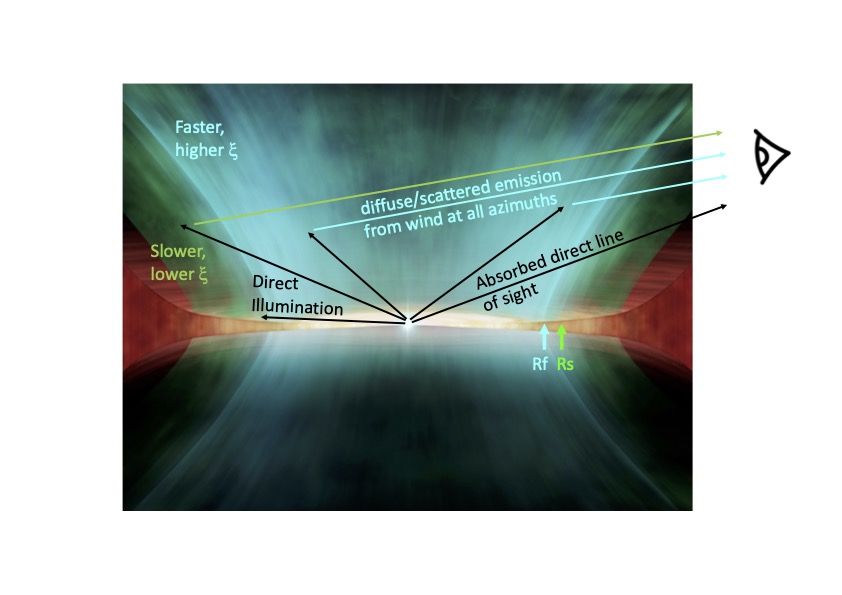}\\
\vspace{-12mm}
\begin{spacing}{1.0}
\caption{\textbf{\label{fig:sketch} Artist's impression of the wind in \gx\ as seen by \xrism.}
The bulk of the wind (green) is optically thick, highly ionised and slow but it has a 
faster, even more highly ionised skin on its inner edge (blue). We see the central source directly through this heavy absorption, but the irradiated wind material forms a secondary source of diffuse X-rays from scattering and re-emission which can be seen along multiple paths. Credit: CfA/Melissa Weiss}
\end{spacing}
\end{figure}

Our final model incorporates all these components. We use two absorption zones: a slow absorber to match the bulk of the material, and a faster column to match the blue wing on the most highly ionised lines. 
The model also includes the line and recombination emission calculated by the photoionision code
for these two absorption columns to approximate the reprocessed emission from the wind. 
These emission components are assumed to be at rest but velocity broadened, as expected 
for emission over all azimuths, and are partially absorbed to approximate the multiple sightlines (see Figure~\ref{fig:sketch}). We also include an additional diffuse component from electron scattering of the intrinsic continuum. 
This is the model (orange) shown with the data in Figure~\ref{fig:resolve-label}, and Figure~\ref{fig:resolve-zoom}, with full fit parameters in 
Methods (Extended Data Table~\ref{tab:2pion_2emit_XR}).

The faster wind column density is difficult to robustly constrain. Most of the elements are fully ionised so there is a large and model-dependent correction between the observed blue wing and the total column. For example, at our best fit ($\log\xi=4.69\pm 0.01$), only 10\% of Fe is visible as Fe~{\sc XXVI}; all remaining Fe is completely ionised (see Methods, Extended Data Figure~\ref{fig:fe_frac}). The fast wind in our best fit model has a column density of 
$N_{\rm H}=(0.79\pm0.09)\times 10^{24}$~cm$^{-2}$, increasing the line-of-sight optical depth to $\tau_{es}\approx 1.8$. Correcting for this attenuation results in  an even higher estimate for the intrinsic luminosity of $L\approx 1.8L_{\rm Edd}$, giving the upper limit of the band of likely intrinsic fluxes shown in blue in Figure~\ref{fig:archive}. 

The two wind components are most likely an approximation of a continuous wind structure as they have similar kinematics. The inner face of the wind (smallest radii) is more highly illuminated and faster, slightly shielding the less ionised, slower material at larger radii. This assumed geometry allows us to estimate the physical parameters of the outflow (Methods). 
The wind is launched from $R_{\rm f} \approx 3\times 10^4 R_{\rm g}$ ($6\times 10^9$~cm), with initial density $\approx 10^{14}$ cm$^{-3}$ for a source of intrinsic luminosity $1.8L_{\rm Edd}$. Attenuation by electron scattering reduces the flux to approximately $ L_{\rm Edd}$ by $R_s\approx 7\times 10^4 R_{\rm g}$
($1.5\times 10^{10}$~cm), beyond which the wind is slower. This is shown schematically in Figure~\ref{fig:sketch}.

We can estimate the mass loss rate of the wind if we can independently estimate the solid angle, $\Omega$ subtended by the wind. For optically thin winds this can be determined from the contribution of scattered emission to the total flux as $f_{\rm scatt}\approx (\Omega/4\pi)(1-e^{-\tau_{es}})$. The observed scattered fraction is difficult to robustly constrain as it depends on the details of how the diffuse flux is modelled. In our fits it ranges from 
$0.22-0.04$ ( Methods and Extended Data Tables \ref{tab:ion_ion} and  \ref{tab:2pion_2emit_XR}).
We assume that these numbers bracket the true scattered flux, so $0.04<(\Omega/4\pi)<0.22$. The assumed mass profile gives a larger uncertainty, and both together 
give a range in total (fast plus slow) mass loss rate of $1.2<\dot{M}_{18}<39$, where $\dot{M}_{18}$ denotes units of $10^{18}$~g s$^{-1}$ (Methods). This is approximately $0.3-10\times$ the inferred mass accretion rate onto the neutron star. This highly non-conservative mass transfer, with as much or more mass being ejected from the system than is accreting, is often seen in galactic binary disc winds\cite{Neilsen2011,Tomaru2020}. Nonetheless, the kinetic power in this wind is very much smaller than the radiative power, as its velocity (even with the faster component) is much less than $ c$.

Similarly high column winds were suggested 
to explain rare observations of black hole binaries
with potentially similar properties (near-Eddington or super-Eddington flux, large disc, high inclination \cite{Miller2015,Neilsen2016,Miller2020,Neilsen2020}). However, 
without both (1) broadband spectra to show the edge depth as in Figure~\ref{fig:archive} and (2) sensitive high-resolution spectra above 7 keV to reveal unsaturated high-order lines (Figure~\ref{fig:resolve-label}, bottom), it is difficult to distinguish between an intrinsically dim source with an optically thin wind ($N_{\rm H}\approx {\rm few}\, \times 10^{23}$~cm$^{-2}$) and a source that is much brighter but strongly attenuated by an optically thick wind.

Accretion disc winds in X-ray binaries
are often viewed as small-scale versions of the winds from supermassive black holes that drive much of AGN feedback: whether these winds are launched by magnetic fields, radiation pressure, or Compton heating remains an open question across the mass scale\cite{Ponti2012, DiazTrigo2016,Neilsen2023,Parra2024}.
The \xrism\ observation of \gx\ provides one of the most sensitive probes of the physics of accretion disc winds to date. 
We therefore consider the origin of this wind by comparing with expectations for winds from all of these driving mechanisms. 

The source is at or above the Eddington limit, meaning that radiation from within the disc is strong enough to directly launch its own photosphere as a wind.
For $L\approx  1-2L_{\rm Edd}$ this occurs only in the inner disc, in which the local flux peaks. This Eddington wind should be fast, with mildly relativistic velocities $v_{\rm out}\approx 0.2\,c$ (ref. \citenum{Ohsuga2011}), not at all compatible with the observed wind in \gx. 

Instead, radiation pressure from the central source could launch a wind by illuminating material at any other radius in the disc, as the effective gravity is proportional to $(1-L/L_{\rm Edd})$ (refs. \citenum{Proga2002,Done2018}). However, this illumination also heats the disc surface to the radiation temperature. In a sub-Eddington source this heating alone can be sufficient to drive a wind from radii where the sound speed exceeds the local escape velocity (called thermal or Compton heated winds)\cite{Begelman1983a,Woods1996,Done2018}. Radiation pressure acts as a boost to the velocity for $L>0.2L_{\rm Edd}$ (thermal-radiative winds)\cite{Tomaru2019,Higginbottom2017}.

Thermal-radiative driving can give a fairly 
good match to most previous data on binary 
winds\cite{DiazTrigo2016, Allen2018, Hori2018, 
Shidatsu2019}, but in their simplest form these models predict narrow lines, as all 
the material is heated to the same temperature and therefore expands with constant velocity\cite{Tomaru2023a}. However, 
detailed radiation hydrodynamical simulations of thermal-radiative winds from large accretion discs 
show that these winds do start to become stratified at high luminosities ($L=0.5-0.7L_{\rm Edd}$) because of optical depth effects.
These more reliastic simulations of thermal-radiative winds have
faster, higher ionisation material on the inner, more strongly illuminated face of the wind, with velocities closely matched to those 
seen in  \gx (refs. \citenum{Tomaru2020,Tomaru2023a,Higginbottom2018}) (see also refs. \citenum{Waters2021,Ganguly2021}). However, the predicted column densities along the line of sight are only a few $\times 10^{23}$~cm$^{-2}$, a factor of 10 below those required here. Part of this discrepancy is probably because that the simulations do not extend to super-Eddington luminosities, but
an additional problem is that
current codes do not yet include scattered emission from the wind
in calculating the illumination of the disc to launch the wind. This scattered flux can exceed the direct illumination when the wind becomes optically thick\cite{Tomaru2023b}.

Alternatively, magnetic winds can also give 
a stratified velocity/ionisation structure. The drawback of this driving mechanism is that the 
magnetic field configuration (and consequent wind density) cannot now be predicted from first principles, but assuming self-similar, large scale fields  connecting into the disc at all radii
\cite{Blandford1982,Contopoulos1994}, gives a wind density structure $n(R)\propto R^{-p}$ 
with $p=1-1.5$ (refs. \citenum{Fukumura2017,Datta2024}). This
predicts that material launched at smaller radii has faster velocity and higher ionisation state, 
generically producing blueward asymmetric line profiles, as observed in the FeXXVI Ly$\alpha_1$ line (see Figure~\ref{fig:resolve-zoom}).
However, this self-similar model is problematic for \gx\ as it predicts 
an additional optical depth of $\tau_{\rm es}\approx 6$ in fully-ionised material inwards of the fast component. This would require an uncomfortably large intrinsic source luminosity to produce the observed X-ray flux. 

Whatever the physical origin, the observed slow wind 
can be used to put upper limits on the kinetic power of any fast wind from the inner disc along this line of sight, such as a
radiation pressure driven wind
from the inner, bright $L\gtrsim L_{\rm Edd}$ flow. These winds
have predicted velocities
$v\sim 0.1-0.2c$ and kinetic power of $\sim 0.05L_{\rm Edd}$ in both analytic and numerical simulations
\cite{Poutanen2007,Sadowski2016,Jiang2024}, but such high ram pressure material along our sight line would strongly disrupt the slow, quiet kinematics of the observed outer disc wind. Our data require that any fast, wind produced by the 
$L\approx L_{\rm Edd}$ inner disc regions must be collimated 
in the polar direction, potentially 
by the formation of a inner disc funnel. 

This is important as fast, inner 
disc winds with $v\sim 0.2c$ are seen in supermassive black 
holes, most compellingly from those 
with $L\gtrsim L_{\rm Edd}$ (for example, 
PDS456; ref. \citenum{Nardini2015}). This 
AGN wind was confirmed 
by recent \xrism\ data, where the 
strong emission as well as 
absorption signatures require that 
the wind is quasi-spherical\cite{xrism2025}, unlike any inner disc wind in \gx. 
Understanding this difference in inner disc and wind properties across the mass scale will lead to a deeper understanding of the physics of AGN feedback across cosmic time. 


\setlength{\bibsep}{0.0pt}
\smallskip
{\sf\small

}

\begin{addendum}

\item[XRISM collaboration]

Marc Audard$^{1}$,
Hisamitsu Awaki$^{2}$,
Ralf Ballhausen$^{3,4,5}$,
Aya Bamba$^{6}$,
Ehud Behar$^{7,27}$,
Rozenn Boissay-Malaquin$^{8,4,5}$,
Laura Brenneman$^{9}$,
Gregory V.\ Brown$^{10}$,
Lia Corrales$^{11}$,
Elisa Costantini$^{12}$
% ...
\\

\begin{affiliations}

  \item Department of Astronomy, University of Geneva, Versoix CH-1290, Switzerland.
  \item Department of Physics, Ehime University, Ehime 790-8577, Japan.
  \item Department of Astronomy, University of Maryland, College Park, MD 20742, USA.
  \item NASA / Goddard Space Flight Center, Greenbelt, MD 20771, USA.
  \item Center for Research and Exploration in Space Science and Technology, NASA / GSFC (CRESST II), Greenbelt, MD 20771, USA.
  \item Department of Physics, University of Tokyo, Tokyo 113-0033, Japan.
  \item Department of Physics, Technion, Technion City, Haifa 3200003, Israel.
  \item Center for Space Science and Technology, University of Maryland, Baltimore County (UMBC), Baltimore, MD 21250, USA.
  \item Center for Astrophysics | Harvard-Smithsonian, MA 02138, USA.
  \item Lawrence Livermore National Laboratory, CA 94550, USA.
  \item Department of Astronomy, University of Michigan, MI 48109, USA.
  \item SRON Netherlands Institute for Space Research, Leiden, The Netherlands.
  \item ESO, Karl-Schwarzschild-Strasse 2, 85748, Garching bei M\"unchen, Germany.
  \item Centre for Extragalactic Astronomy, Department of Physics, University of Durham, South Road, Durham DH1 3LE, UK.
  \item Institute of Space and Astronautical Science (ISAS), Japan Aerospace Exploration Agency (JAXA), Kanagawa 252-5210, Japan.
  \item Department of Physics, Kyoto University, Kyoto 606-8502, Japan.
  \item Department of Economics, Kumamoto Gakuen University, Kumamoto 862-8680, Japan.
  \item Department of Physics, Tokyo Metropolitan University, Tokyo 192-0397, Japan.
  \item Department of Physics, Hiroshima University, Hiroshima 739-8526, Japan.
  \item Department of Physics, Fujita Health University, Aichi 470-1192, Japan.
  \item Department of Astronomy and Physics, Saint Mary's University, Nova Scotia B3H 3C3, Canada.
  \item Cahill Center for Astronomy and Astrophysics, California Institute of Technology, Pasadena, CA 91125, USA.
  \item European Space Agency (ESA), European Space Research and Technology Centre (ESTEC), 2200 AG, Noordwijk, The Netherlands.
  \item Faculty of Engineering, University of Miyazaki, Miyazaki 889-2192, Japan.
  \item RIKEN Nishina Center, Saitama 351-0198, Japan.
  \item Leiden Observatory, University of Leiden, P.O. Box 9513, NL-2300 RA, Leiden, The Netherlands.
  \item Kavli Institute for Astrophysics and Space Research, Massachusetts Institute of Technology, MA 02139, USA.
  \item Department of Physics, Saitama University, Saitama 338-8570, Japan.
  \item Department of Physics, Rikkyo University, Tokyo 171-8501, Japan.
  \item Faculty of Physics, Tokyo University of Science, Tokyo 162-8601, Japan.
  \item Faculty of Science and Technology, Tokyo University of Science, Chiba 278-8510, Japan.
  \item Department of Electronic Information Systems, Shibaura Institute of Technology, Saitama 337-8570, Japan.
  \item Department of Earth and Space Science, Osaka University, Osaka 560-0043, Japan.
  \item Department of Physics, University of Wisconsin, WI 53706, USA.
  \item Department of Physics and Astronomy, University of Waterloo, Ontario N2L 3G1, Canada.
  \item Department of Physics, Nagoya University, Aichi 464-8602, Japan.
  \item Science Research Education Unit, University of Teacher Education Fukuoka, Fukuoka 811-4192, Japan.
  \item Hiroshima Astrophysical Science Center, Hiroshima University, Hiroshima 739-8526, Japan.
  \item Department of Data Science, Tohoku Gakuin University, Miyagi 984-8588.
  \item College of Science and Engineering, Kanto Gakuin University, Kanagawa 236-8501, Japan.
  \item European Space Agency(ESA), European Space Astronomy Centre (ESAC), E-28692 Madrid, Spain.
  \item Department of Science, Faculty of Science and Engineering, KINDAI University, Osaka 577-8502, JAPAN.
  \item Department of Teacher Training and School Education, Nara University of Education, Nara 630-8528, Japan.
  \item Astronomical Institute, Tohoku University, Miyagi 980-8578, Japan.
  \item Department of Physics, Nara Women's University, Nara 630-8506, Japan.
  \item School of Science and Technology, Meiji University, Kanagawa, 214-8571, Japan.
  \item Yale Center for Astronomy and Astrophysics, Yale University, CT 06520-8121, USA.
  \item Department of Physics, Konan University, Hyogo 658-8501, Japan.
  \item Graduate School of Science and Engineering, Kagoshima University, Kagoshima, 890-8580, Japan.
  \item Department of Physics, Chuo University, Tokyo 112-8551, Japan.
  \item Faculty of Education, Shizuoka University, Shizuoka 422-8529, Japan.
  \item Department of Astronomy, Kyoto University, Kyoto 606-8502, Japan.
  \item Nihon Fukushi University, Shizuoka 422-8529, Japan.
  \item Anton Pannekoek Institute, the University of Amsterdam, Postbus 942491090 GE Amsterdam, The Netherlands.
  \item RIKEN Cluster for Pioneering Research, Saitama 351-0198, Japan.
  \item Johns Hopkins University, MD 21218, USA.
  \item Department of Astronomy and Astrophysics, University of Chicago, Chicago, IL 60637, USA.
  \item Villanova University, Mendel Science Center 263B, Villanova PA 19085, USA.
  \item Space Telescope Science Institute, Baltimore, MD 21218, USA.
\item Kavli IPMU (WPI), UTIAS, The University of Tokyo, Kashiwa, Chiba 277-8583, Japan

\end{affiliations}

email: \url{chris.done@durham.ac.uk, r.tomaru.sci@osaka-u.ac.jp, joseph.neilsen@villanova.edu, mizumoto-m@fukuoka-edu.ac.jp}\\

\end{addendum}

\clearpage

\begin{methods}

\section*{Data extraction}\label{sec:data}

\noindent \textbf{\xrism.} The source was observed between 2024-02-25 02:26:51 UT and 2024-02-26 00:06:46
(ObsID 300036010). 
Data reduction was performed with the software
versions of the pre-pipeline version JAXA ``004\_002.15Oct2023\_Build7.011''  and the 
pipeline script ``03.00.011.008'', and the internal CALDB8, which corresponds to the public XRISM CALDB ver.\ 20240815. 

The Resolve data were filtered to exclude periods affected by the Earth's eclipse, the sunlit Earth's limb, South Atlantic Anomaly (SAA) passages, and the initial 4300~s following the recycling of the 50-mK cooler. Events in the resulting good time intervals (GTI) were
screened using pixel-to-pixel coincidence and an energy-dependent rise time cut\citep{Kilbourne2018,mochizuki2025}. Pixels 12 (calibration pixel) and 27 (which shows unexpected gain fluctuations) were excluded. 
The net exposure time after filtering was 37.8~ks, with a total count rate of $72.1~{\rm s}^{-1}$.

A timing coefficient in the CALDB is used to set a flag for any event occurring near-in-time to another event on another pixel. 
The false coincidence for pixel-pixel coincidence may not be ignoreable especially in the bright sources but in our observations, the loss fraction calculated from the 
{\sc status[4]} flag is 
only $\sim10$\%.

Calorimeter events are classified into “grades” based on the time interval from temporally adjacent events. Here we use only 
High-resolution Primary (Hp) grade, which provides the highest energy resolution. The Hp count rate was $30.1~{\rm s}^{-1}$, representing 42\% of the total. A redistribution matrix file (RMF) was generated using \texttt{rslmkrmf} based on the cleaned event file, with a parameter file of xa\_rsl\_rmfparam\_20190101v006.fits. The line-spread function components considered included the Gaussian core, exponential tail to low energy, escape peaks, silicon fluorescence, and electron loss continuum (i.e., the "X" option was selected). An auxiliary response file (ARF) was generated using \texttt{xaarfgen}, assuming a point-like source at the aim point, including the additional opacity of the gate valve closed current configuration of Resolve\citep{Midooka2021}.

\makeatletter
\def\fnum@figure{{\bf Extended Data Fig.~\thefigure}}
\def\fnum@table{{\bf Extended Data Table~\thetable}}
\makeatother
\setcounter{figure}{0}

\begin{figure}[!b]
\centering

\includegraphics[width=\linewidth]{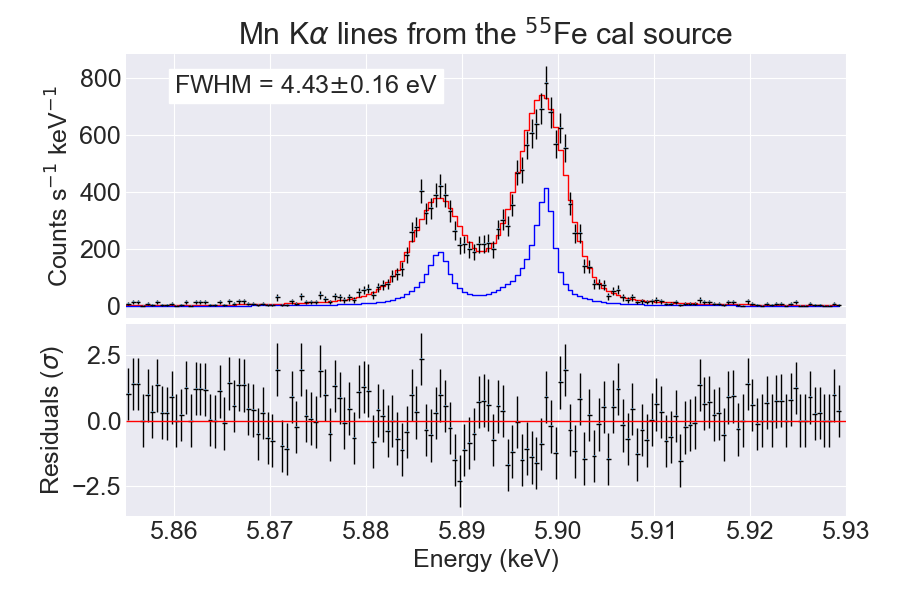}
\begin{spacing}{1.0}
\vspace{-12pt}
\caption{\textbf{\label{fig:Fe55}
Mn K$\alpha$ lines from the $^{55}$Fe source in the filter wheel.} The black bins show the Hp spectrum extracted using two gain fiducial points, summing the 34 pixels. The blue line shows the intrinsic line profile, whereas the red one represents the best fit model, with additional Gaussian broadening of FWHM=4.43~eV. The lower panel shows the residuals between the data and the model, indicating that this is a good description.}
\end{spacing}
\end{figure}

The temperature sensitivity of the Resolve detector necessitates pixel-by-pixel correction for gain drift to maintain proper energy scale and resolution. The gain scale function is parameterized by an “effective temperature” \citep{porter16}, which was tracked over time for each pixel. The Mn K$\alpha$ line complex from the \(^{55}\)Fe calibration source was used to calculate the effective temperature. Two gain fiducial measurements were performed at the start and end of the observation, where the entire array was illuminated by the \(^{55}\)Fe source in the filter wheel. The Mn K$\alpha$ spectrum, shown in Extended Data Figure~\ref{fig:Fe55}, has an energy resolution of $4.43\pm0.16$~eV (FWHM) and an energy offset of less than 0.16~eV. A significant temperature shift was identified after the observation, attributed to spacecraft maneuvers and orientation. In the calibration pixel, continuously illuminated by \(^{55}\)Fe, the effective temperature shifted from 49.969 mK to 49.965 mK and then to 49.975 mK (Extended Data Figure~\ref{fig:gaintrend_calpix}). If the gain drift were tracked only at the fiducial points, the maximum effective temperature shift would be 0.005~mK, corresponding to a 1.5~eV energy shift (Extended Data Figure~\ref{fig:gaintrend_calpix}). 
To reduce this shift, we introduced an ad-hoc gain point (red circle in Extended Data Figure~\ref{fig:gaintrend_calpix}), and calculated the effective temperature difference ($\Delta T_\mathrm{eff}$) between the initial and ad-hoc gain point. Scaling from the gain change on the calibration pixel at this intermediate point, we added a new gain point for each of the other pixels (Extended Data Figure~\ref{fig:gaintrend}) and corrected the X-ray energies using linear interpolation.  After this observation, a gain fiducial 9 hours after a maneuver was added to standard operations.

\begin{figure}
\centering
\includegraphics[width=\linewidth]{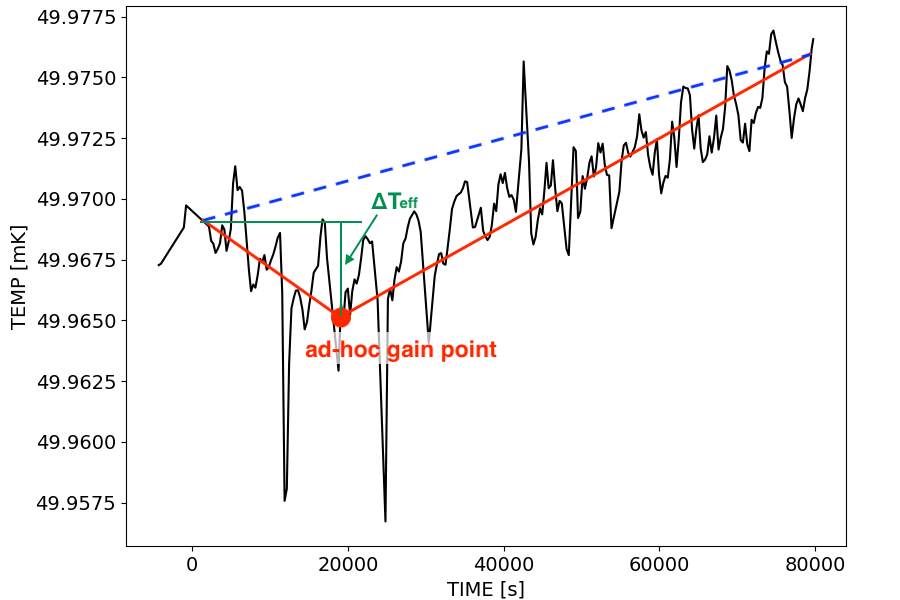}
\begin{spacing}{1.0}
\caption{\textbf{\label{fig:gaintrend_calpix} Effective temperature of the calibration pixel versus time.} The effective temperature across the observation is shown as a solid black line, compared to a linear interpolation between the measurements at the start and end of the observation (blue dashed line). We introduce an ad-hoc gain point (red filled cicle, with a temperature $\Delta T_\mathrm{eff}$ below the first gain point), to give a better match (red solid line). }
\end{spacing}
\end{figure}

\begin{figure}
\centering
\includegraphics[width=\linewidth]{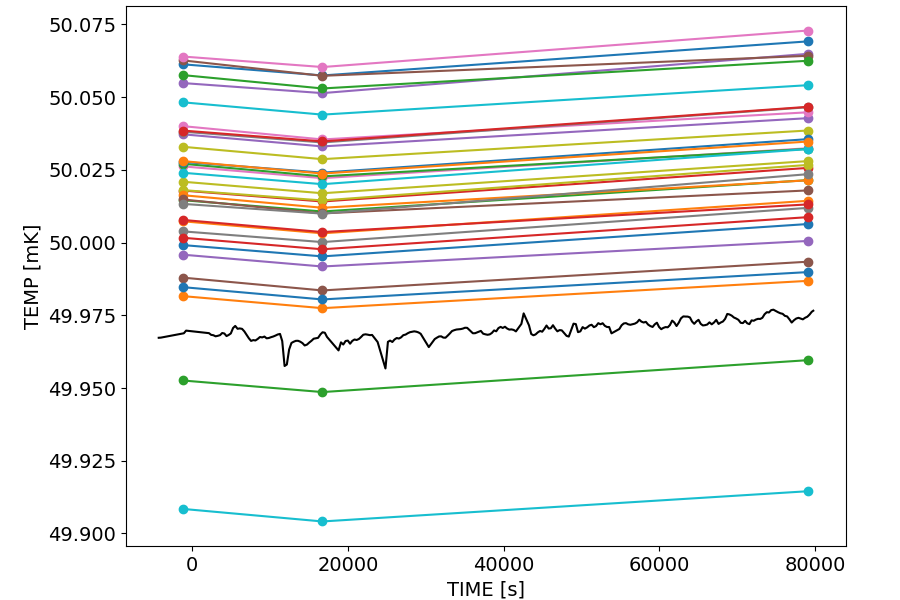}
\begin{spacing}{1.0}
\caption{\textbf{\label{fig:gaintrend}Effective temperature variations in all pixels except 27.} Each pixel has an effective temperature estimate corresponding to the gain fiducial measurements at the beginning and end of the observation. We introduced an additional gain point by scaling the ad-hoc gain point from the calibration pixel (see Extended Data Fig. \ref{fig:gaintrend_calpix}) to each individual pixel (see the middle point in each colored line). The black line shows the calibration pixel, which is tracked continuously, for reference.}
\end{spacing}
\end{figure}

At high count rates, energy resolution degradation may occur due to contamination from untriggered electrical cross-talk events\cite{mizumoto2025}. 
To evaluate this effect we use spectra of Cr Ly$\alpha_1$ and $\alpha_2$ lines from the source, which are strong and close in energy to the Mn K$\alpha$ calibration lines. We compare these before and after cross-talk effect screening, but the line widths do not change significantly. The results indicate that cross-talk has a 
negligible effect on the energy resolution in this observation; therefore, we do not apply cross-talk screening in order to  preserve photon statistics.

We also check whether there is contamination of the data from pseudo-Ls (Low-resolution Secondary) events. However, this is more important for fainter sources, and is negligible here below 10~keV. 

The data are not corrected for any systemic velocity offsets as these are small, with a blueshift of 40~km/s for the combined effects of the velocity of the Earth round Sun and the 
Galactic rotation at \gx\ position with respect to the local standard of rest. The peculiar velocity of the binary system is also fairly small, at  
$-70\pm 30$~km/s\cite{Bandyopadhyay1999}. 


\noindent \textbf{NuSTAR.} 
The \nustar\ observation took place from 2024-02-25 12:56:09 UT to 2024-02-26 10:36:00 UT (ObsID 30901010002).
We use the {\tt nupipeline} from HEASOFT v.6.33.2 to reduce \nustar\ observation 30901010002. We use {\tt nuproducts} to extract \nustar\ source and background spectra and to create response files. The source region is a 1 arcminute circle centered on the source; we used surrounding source-free regions for background.
To maximize simultaneity and mitigate the effects of any source variability, we define the \nustar\ good time interval as the beginning of the \nustar\ observation to the end of the \xrism\ observation.\vspace{-4mm}

\section*{Ion-by-ion model fitting}\label{sec:ionbyion} 

Disc accreting neutron star continuum spectra are generally well modelled by a multicolour disc component, together with higher temperature emission from a boundary layer between the disc and NS surface. The boundary layer illuminates the disc, producing a characteristic reflection spectrum that is broadened by the relativistic velocities and strong gravity of the inner disc. We approximate this component with a broad gaussian  {emission} line  {with energy fixed at 6.4~keV}. The \xrism\ spectra are rebinned to require 20 counts per bin and fit between 2.1 and 18 keV  {using the {\tt xspec}\cite{Arnaud1996} X-ray spectral fitting package.}

Relative to this absorbed disk plus blackbody and gausssian continuum, the fit residuals show numerous narrow absorption features. We first model these features by considering each ion independently. 
 {The {\tt kabs} model\cite{Tomaru2020} (a local model for use in {\tt xspec} software) calculates the full Voigt absorption line profile for a single transition in a given ion. Modelling the full series of transitions from $n=1$ for a given ion then involves multiple {\sc kabs} components, with ion column and velocity outflow and width tied across all the components. We develop a more convienient {\tt xspec} local model, {\tt Ionabs},
which packages all these together, calculating the full line series from a given ion column with given kinematics in a single model component. This includes all fine structure lines, as well as the self-consistent edge structure(s) (including the L-shell edge for ions with 3 or more electrons such as FeXXIV). The Voigt profile velocity width in {\sc kabs} is defined as $v_{turb}/c=(E-E_0)/\Delta E$, but the photo-ionization code {\tt pion} (see below) uses the gaussian width $v_{rms}/c=(E-E_0)/(\sqrt{2}\Delta E)$ so here we report $v_{rms}$ so that these can be directly compared}. Transition energies, oscillator strengths, Einstein A values and the energy dependence of the cross sections are taken from Flexible Atomic Code\cite{Gu2008}. These match very well to
the NIST database for H- and He-like ions. 

It is immediately clear that most of the lines have similar kinematic structure with slow outflow velocity and very narrow
velocity width. This "slow" wind component must have very high column density in order to produce the 
multiple higher order lines (transitions beyond $n=8$) of Fe\,{\sc xxv} and Fe\,{\sc xxvi}. Such a column should have corresponding $n=2-1$ K$\alpha$ transitions which are completely saturated so the lines cores should be completely black. The fact that the data never 
go to zero shows that there is an additional 
diffuse source of X-rays, most likely from the wind itself (see Figure~\ref{fig:resolve-zoom}). Additionally, the
detailed Fe\,{\sc xxvi} K$\alpha_1,\alpha_2$ line profile 
clearly shows a strong blue wing, requiring that there is an additional, faster wind component present (see Figure~\ref{fig:resolve-zoom}).

Thus we model the intrinsic spectrum absorbed by two wind components, "slow" (16 ions) and "fast" (only Fe\,{\sc xxv}, Fe\,{\sc xxvi}, Ni\,{\sc xxvii}, Ni\,{\sc xxviii}). 
Modelling the diffuse emission is more challenging as it should be extremely complex, with recombination radiation from the X-ray heated material, and scattered incident continuum forming a spatially extended source which is absorbed along multiple different lines of sight through the wind. We first
approximate this as electron scattering alone, so a fraction 
$f_{scatt}$ of the incident continuum,
but a better fit to the remaining residuals around the absorption lines is if the scattered continuum is also absorbed by the "fast" wind component.


The model then consists of the intrinsic continuum
${\tt Int=(diskbb+gauss+bbody)},$
absorbed by multiple ions grouped into two kinematic components (slow: Ionabs$_{\rm s}$ and fast: Ionabs$_{\rm f}$), together with photoelectric absorption from neutral material ({\tt TBabs}) fixed at the interstellar column density of $N_{\rm H}=3.2\times10^{22}$ cm$^{-2}$. The model is\\
${\tt TBabs\times (Ionabs^{16}_s Ionabs^{4}_f + f_{scatt} Ionabs^{4}_f)\times Int}$ in {\sc xspec} notation, where the superscripts on  {\tt Ionabs} show the number of ions included. 

The resulting fit parameters are given in Extended Data Table \ref{tab:ion_ion}. The ions in the slow component all have 
similar outflow velocities of $v_{\rm out} \sim 330~{\rm km/s}$.
The fast component appears to have a wider range of kinematics, with outflow velocities ranging from $\sim$ 500 to $\sim$1000 km 
s$^{-1}$ depending on the ion, but both fast and slow components have line widths $\sim v_{out}/2$. 

The column densities derived for the ions in the slow component (Extended Data Table~\ref{tab:ion_ion}) are almost an order of magnitude larger than found in previous {\it Chandra}/HETGS data (ref.\ \cite{Ueda2004}). We estimate  {a lower limit to} the equivalent Hydrogen column density from adding the slow component Fe~{\sc XXV} and Fe~{\sc XXVI} ion columns together, to get $N_{\rm Fe}=46^{+6}_{-5}\times 10^{18}$~cm$^{-2}$, giving $N_{\rm H}>N_{\rm Fe}/A_{\rm Fe}=1.4\pm 0.2\times 10^{24}$~cm$^{-2}$ for $A_{\rm Fe}=3.3\times 10^{-5}$.

This column density is large enough that electron scattering optical depth is significant:
$\tau_{\rm es}=1.21\,N_{\rm H}\,\sigma_{\rm T}\gtrsim 1.1$, where the factor 1.21 comes from the number of electrons per Hydrogen atom in a fully-ionised plasma of solar abundance, and $\sigma_{\rm T}$ is the Thomson cross-section. The observed Fe ion columns in the slow wind component already imply that the wind is optically thick to electron scattering, and yet there should be even more material, firstly due to the fast wind and secondly since some fraction of Fe is completely stripped to Fe~{\sc XXVII} so produces no line signatures. 
This correction need not be very large for 
the slow wind, where the ratio of columns 
Fe~{\sc XXVI}/Fe~{\sc XXV} is close unity. However, this is not true for the fast wind, where the column in Fe~{\sc XXVI} is $2.5\times$ larger than that of Fe~{\sc XXV}.
Thus the observed ion columns in the fast wind are likely only a small tracer of the likely column present. To correct for this requires an ionisation balance calculation.

\section*{Photoionisation modeling for ion fractions}\label{sec:pion} We use the photoionised plasma model {\tt pion} version 3.08.00, available in {\sc spex}\cite{Kaastra1996,Mehdipour15}. We compute the ratios of different ionisation stages $N^{i+1}/N^{i}$ (Extended Data Figure~\ref{fig:ion_f}) and the ion fractions (Extended Data Figure~\ref{fig:fe_frac}) as functions of the ionisation parameter $\log \xi\equiv L/(nR^2)$. We assume an intrinsic illuminating continuum which matches the best fit incident continuum (disk blackbody plus blackbody). 


All of ion ratios in the slow component are consistent with that from Fe\,{\sc xxvi}/Fe\, {\sc xxv} alone, giving an ionisation parameter of $\log\xi= 3.85-3.98$. 
This gives an ion fraction for completely ionised iron (Fe\,{\sc xxvii}) of $f_{\rm xxvii} = 0.14-0.23$ (Extended Data Figure~\ref{fig:fe_frac}). This increases the total iron column density in the slow wind  to 
$N_{\rm Fe} = 51-60\times10^{18}$ cm$^{-2}$, leading to an equivalent hydrogen column density $N_{\rm H} =N_{\rm Fe}/A_{\rm Fe} =1.5-1.8 \times 10^{24}$cm$^{-2}$ 
and $\tau_{es}=1.2-1.4$.

A similar analysis for the fast wind gives a more complex picture. The Fe ratio suggests $\log \xi =4.15-4.29$
(corresponding to ion fraction of Fe~{\sc XXVII} 0.38-0.53)
giving 
$N_{\rm H}\sim 4.3-5.6\times 10^{22}$ cm$^{-2}$. 
However, these broader lines are less well defined in the data, and therefore more sensitive to the model assumed to approximate the complex diffuse emission from the wind. Thus the column density of the faster component is much more uncertain (see the full photoionisation fits below). 

The unabsorbed continuum model (without scattered flux) gives a bolometric flux of $F = 9.1\times 10^{-9} ~{\rm erg~cm^{-2}~s^{-1}}$ (13.6 eV--100 keV). Correcting this for attenuation by electron scattering with $\tau_{es}=1.2-1.4$ gives an intrinsic flux of 
$F_{0}=3.0-3.8\times 10^{-8} ~{\rm erg~cm^{-2}~s^{-1}}$.
The luminosity of this source $L = 4\pi d^2 F_{0}=0.8-1~L_{\rm Edd}$, where $d = 7~{\rm kpc}$ and $L_{\rm Edd} =2.1\times 10^{38}$ erg s$^{-1}$. 

The scattered fraction parameter in these fits $f_{\rm scatt}$ is the ratio of scattered to observed direct flux. We use this to calculate the ratio of scattered to intrinsic flux $F_{\rm scatt}/F_0 = 0.052-0.066$ and use this to estimate the solid angle of the wind, as $F_{\rm scatt}/F_0 \approx  \Omega/4\pi (1-\exp(-\tau_{es}))$. This gives $\Omega/4\pi = 0.08\pm0.01$, though again this is quite uncertain as it depends on the detailed wind geometry and emission/absorption. 

\section*{Fitting with Photoionisation models }\label{sec:fits} We now use the same photoionised code, {\sc pion}  to 
directly fit to the data. We 
calculate a grid of models for solar abundances, simulating absorbers with different values of column density $N_H$, ionisation parameter $\log\xi$ and turbulent velocity $v_{\rm rms}$, fixing the illuminating SED shape to that derived from spectral fitting. Each of the simulations has 65536 logarithmically-spaced bins to cover the energy range from $10^{-4}$ to $10^{3}$ keV with a resolution of 1.5 eV around 6 keV, enough to fit {\it Resolve} data. In total we perform 8736 simulations with values $21\leq \log N_{\rm H} \leq 25$ spaced by 0.2 (21 points), $2 \leq \log \xi \leq 7$ spaced by 0.2 (26 points), and $-5\leq \log (v_{\rm rms}/c) \leq  -2$ spaced by 0.2 (16 grid points). We fit with a single number density $n_{p} = 10^{14}$ cm$^{-3}$ to reduce the size of the tables (see Main text and below); we calculate the population levels from radiative recombination, cascade, radiative and collisional excitation correctly for meta--stable levels\cite{Tomaru2023a, Mao2017}. We build these results into a multiplicative absorption table model\cite{Tomaru2023b} for use in {\sc xspec}.

\begin{figure}
\centering
\includegraphics[width=0.9\linewidth]{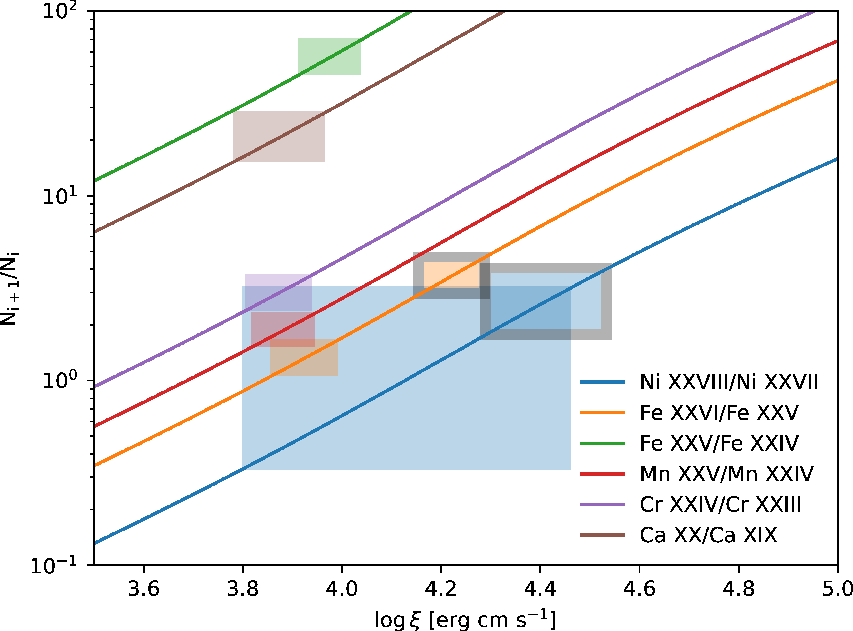}
\begin{spacing}{1.0}
\caption{\textbf{\label{fig:ion_f} Ion ratio as a function of ionisation parameter.} We computed the ground state populations for each ion using the {\tt pion} code as in Methods.
The ratio of these populations (equivalently, the ratio of the column densities in different ions) is sensitive to the ionisation parameter, as shown. 
Using the ratio of column densities taken from Extended Data Table \ref{tab:ion_ion}, we estimate the ionisation parameter of the slow component in our ion-by-ion fits as $\log\xi\sim3.9$(shaded regions), 
and the fast component of Fe and Ni as $\log\xi=4.15-4.53$ (shaded regions with black frames).} 
\end{spacing}
\end{figure}

\begin{figure}
\centering
\includegraphics[width=0.9\linewidth]{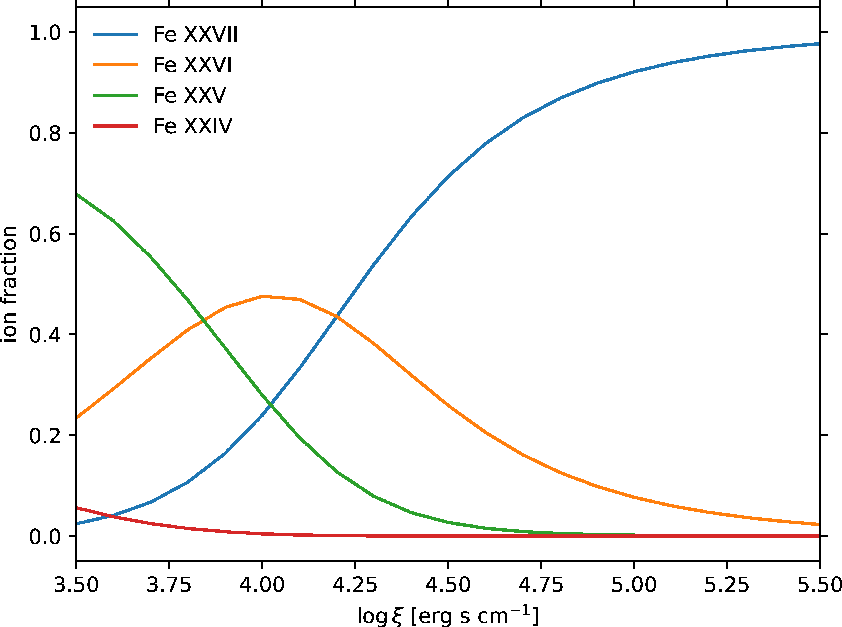}
\begin{spacing}{1.0}
\caption{\textbf{\label{fig:fe_frac} Ion fractions of Fe versus the ionisation parameter}. This is computed using {\tt pion} as described in Methods, 
assuming that the gas is photoionised by the
continuum shape observed. 
We estimate the ionisation parameter from our ion-by-ion fits using Extended Data Figure \ref{fig:ion_f}, then used the curves above to determine the column density of completely-ionised Iron (Fe\,{\sc xxvii}).}

\end{spacing}
\end{figure}


\noindent \textbf{Diffuse continuum approximated by absorbed scattered flux}
We replace the multiple ion-by-ion absorption components with two {\tt pion} absorption components (one fast: ${\tt abs_f}$ and one slow: ${\tt abs_s}$). As above, we assume that the diffuse emission has the same shape as the incident continuum 
and that this scattered component is absorbed by the fast wind. 
We represent this model in {\sc xspec} form as ${\tt TBabs}\times ({\tt abs_{s}} {\tt abs_{f} Int} +  f_{\rm scatt} {\tt abs_{f} Int )}$. 

Remarkably, the goodness of the fit is not far from that of the “ion-by-ion” fit, which allowed  {free element abundances and allowed} every ion to have different kinematics and ion ratios ($\chi2=14726/13555$ i.e. 52 fewer free parameters, for $\Delta\chi^2\sim 550$). This gives very similar plasma parameters for the slow component as derived from the ion-by-ion fitting, namely $\log \xi = 3.93^{+0.01}_{-0.02}$ and $N_{\rm H} =154^{+8}_{-6}\times 10^{22}$ cm$^{-2}$. 
However, the fast component now has a higher ionisation parameter ($\log\xi=4.49$) and hence a 
higher equivalent column density of $N_{\rm H}
=9^{+2}_{-1}\times 10^{22}$ cm$^{-2}$. This shows that the fast component parameters are more sensitive to the details of the model, but that the slow component is very robust, and robustly gives an optical depth $\tau_{es}>1$.

\smallskip

\noindent \textbf{Diffuse component including line/recombination emission.} A better approximation to the diffuse/scattered continuum requires using {\tt pion} to 
calculate the emitted line/recombination continuua from the photoionised material in addition to the absorption lines. We take the same incident spectrum and density as for the absorption table model to generate an additive table model for {\sc xspec} (hereafter ${\tt emm}$), but this time we set the solid angle  fraction $\Omega/4\pi = 1.0$.  In principle, the resulting emission normalisations allow the solid angle to be independently estimated, but these will be dependent on details of the radiation transfer through the optically thick wind, so we do not use them here. 

We tie the ionisation parameter and column density to be the same for the absorbing and emitting plasma. We expect the emission should arise from all azimuths, so we fix the outflow velocity to zero, and allow the broadening to be free. We allow this to be somewhat self absorbed by the wind in our line of sight, but caution that this is just an approximation to a more complex geometry that requires a full radiation transfer calcluation. 

The final model is 
$\tt TBabs\times (abs_{s}abs_{f} Int +  
f_{\rm scatt} abs_{\rm f} Int + abs_{\rm s} emm_{\rm f} +$
$\tt f_{\rm scatt} emm_f + emm_{\rm s} )$, where again we fix {\tt tbabs} 
to the interstellar column density of $N_{\rm H}=3.2\times10^{22}$ cm$^{-2}$. 
This gives our best description of the spectrum (Extended Data Table \ref{tab:2pion_2emit_XR}), and is the model shown in Figure~\ref{fig:resolve-label}. This gives a goodness of fit of
$14339/13551$. This has 4 more free parameters than the previous description of the diffuse flux, but gives $\Delta\chi^2=-386$. It is now statistically equivalent to the ion-by-ion fits in Extended Data Table \ref{tab:ion_ion}, as it 
has $48$ fewer free parameters for an increase of $\Delta\chi^2=+151$. 

The emission lines are not obvious by eye as in this model they are dominated by the fast component and are therefore moderately broadened. Nonetheless, the resultant P Cygni profile can be 
seen in FeXXVI Ly$\alpha_{1,2}$ at around 6.94~keV as shown in  
Figure~\ref{fig:resolve-zoom}. The emission lines 
also contribute to the shape of the saturated line cores, and this more complex model
shifts the ionisation parameter of the fast wind to even higher values, $\log\xi=4.69^{+0.03}_{-0.04}$ requiring even higher column densities: $N_{\rm H}=80\times 10^{22}$ cm$^{-2}$. Again, the parameters of the slow wind are mostly consistent with previous models, with just slightly lower column ($N_{\rm H}=132^{+7}_{-8}\times 10^{22}$~cm$^{-2}$) and ionisation state ($\log\xi=3.88\pm 0.01$), but similar kinematics, and scattered fraction.

We also perform fits with the {\sc xstar} model {\tt warmabs}\cite{Kallman2009} to explore the overall robustness of our photoionisation analysis. For {\tt warmabs}, we calculate electron level populations using the best fit continuum from fits to a model consisting of an absorbed disc plus {\tt nthcomp}\cite{Zdziarski1996,Zycki1999} with Gaussian lines and a smeared edge near 8 keV. Fits to the Resolve spectrum with {\tt warmabs} were qualitatively and quantitatively very similar to the {\tt pion} fits, requiring a high-column component with a smaller turbulent line width and blueshift and a more highly-ionised, broader, and faster component. These fits will be presented in detail elsewhere, but we note that despite different assumptions about the ionising continuum, radiative transfer, absorption/emission geometry, and the different codes, we still find a total equivalent of the slow wind column density in excess of $1.4\times10^{24}$ cm$^{-2}$.

\section*{Wind Geometry} We assume that the wind is a
continuous structure so
the outer edge of the fast wind must coincide with the inner edge 
of the slow wind\cite{Neilsen2020}. In other words, the column 
density of the fast and slow winds must be given by $$N_{\rm 
H,f}=\int_{R_{\rm f}}^{R_{\rm s}}n(R)dR,\ \ \ \ \ 
N_{\rm 
H,s}=\int_{R_{\rm s}}^{R_{\rm out}}n(R)dR,
$$ where $R_{\rm s}=R_{\rm 
f}+\Delta R_{\rm f}$, $\Delta R_{\rm f}$ is the width of the fast 
wind, and $n(r)$ is the density profile of the wind. But the 
relative locations of the fast and slow winds are also set by 
their relative ionisation parameters. If the ionisation parameter 
at the inner edge of the fast wind is $\xi_{\rm f}=L_0/n_{\rm 
f}R_{\rm f}^2$, then the ionisation parameter at the inner edge of 
the slow wind must be $\xi_s=L_0 \exp(-\tau_{\rm f})/n_{\rm s} 
R_{\rm s}^2$. Here $n_{\rm f}$ and $n_{\rm s}$ are the densities of 
the wind at $R_{\rm f}$ and $R_{\rm s},$ respectively, while $R_{\rm out}$ is the outermost radius at which the wind is produced (which need not be the same as the disc outer radius). The factor 
$\exp(-\tau_{\rm f})$ is an approximation of the attenuation of 
the radiation field by the fast wind. This is appropriate for a relatively small solid angle wind, as inferred here.

For a self-consistent solution, the radius of the slow wind as 
inferred from the column density of the fast wind must match the 
radius implied by the relative ionisation of the two zones. This gives 4 independent constraints, so we can solve for (at most) 4 independent parameters. We first assume a constant density wind, which gives $n_f=n_s=1.6\times 10^{14}$~cm$^{-3}$, for $R_f=4.7\times 10^9$~cm, $R_s=1.0\times 10^{10}$~cm, and $R_{\rm out}=1.8\times 10^{10}$~cm. Alternatively, we assume a power law density distribtion $n=n_{\rm f}(R/R_{\rm f})^{-x}$
for $R_{\rm f}<R<R_{\rm out}=10^{12}$~cm which is the outer disc radius. This has $n_{\rm f}=0.9\times 10^{14}$~cm$^{-3}$ with $x=1.1$ for $R_{\rm f}=6.3\times 10^9$~cm, $R_{\rm s}=3.7\times 10^{10}$~cm
(giving $n_{\rm s}=n(R_{\rm s})=1.3\times 10^{13}$~cm$^{-3}$). 

These densities are very high but the predominantly H-- and He--like ions seen have no density diagostic potential. Instead, previous work on the black hole binary
GRO\,J1655--40, which 
also showed evidence for a Compton thick wind from a likely super-Eddington state\cite{Neilsen2016,Shidatsu2016,Tomaru2023a, Keshet2024}, measured density directly from a meta--stable L-shell absorption line of B-like Fe\,{\sc xxii}\cite{Miller2008, Mitrani2023,Tomaru2023a}. This line transition at $\sim1$\,keV is outside of the current Resolve bandpass, and would likely not be present in the higher ionisation state seen in the \gx\ outflow. However, weak meta--stable lines from K-shell Be-like Fe\,{\sc xxiii} around 6.6\,keV may be used to probe the density\cite{Mao2017} in future modeling. 

In principle, a thermal wind may be launched from all radii $R\gtrsim 0.2 R_{\rm IC},$ where $R_{\rm IC}$ is the Compton radius\cite{Woods1996}. For \gx, this nominal limit is approximately $3.3\times10^{10}$~cm, a factor of a few larger than the inferred launch radius of the fast wind. However, thermal-radiative winds can be expected from much smaller radii when the luminosity approaches the Eddington limit\cite{Done2018}, so our radii are likely consistent with a thermal-radiative wind.

Finally, we calculate the mass loss rate in the wind. Here the wind is being launched from all radii on the disc from $R_{\rm f}-R_{\rm out}$, so we cannot use the standard mass continuity expressions as the wind mass is increasing over this range. Instead, we calculate the total wind mass in this region, $M$, and the time, $t$, it takes to  expand out of this region as
$$M=\int_{R_{\rm f}}^{R_{\rm out}} 4\pi R^2 (\Omega/4\pi) n(R) m dR$$
$$t=\int_{R_{\rm f}}^{R_{\rm out}}\frac{dR}{v(R)}= \frac{R_{\rm s}-R_{\rm f}}{v_f}+\frac{R_{\rm out}-R_{\rm s}}{v_{\rm s}}$$
where $m=2.4\times10^{-24}$~g is the average atomic mass per Hydrogen atom in a cosmic gas, $\Omega$ is the solid angle of the wind 
and $v(R)$ is the radial velocity profile. 

For the constant density wind, these give $\dot{M}=M/t=2.4-6.6\times 10^{18}$~g/s for the solid angles discussed in the Main text ($0.08\leq \Omega/4\pi \leq 0.22$). This is very similar to the estimates given by 
mass continuity $\dot{M}=4\pi R^2 n(R) v(R) m (\Omega/4\pi)$ which can be rewritten for the fast wind as $\dot{M}_f=4\pi m (\Omega/4\pi) (L/\xi_f )=
0.5-1.5\times 10^{18}$~g/s
while the slow wind gives 
$\dot{M}_f=4\pi m (\Omega/4\pi) L\exp(-\tau_f)/\xi_s
=1.3-3.7\times 10^{18}$~g/s.
However, the power law density profile with $x=1.1$
has much more mass at larger radii, so gives much larger 
$\dot{M}=M/t=14-39\times 10^{18}$~g/s.

Even the lowest estimates for the mass loss rate from constant density assumptions
are comparable with the 
central mass accretion rate of $3.9\times 10^{18}$~g/s required to power the inferred X-ray luminosity, while the largest estimates have up to $10\times$ more matter ejected than is accreted.  

\begin{addendum}
\item[Data availability] The \xrism\ Resolve data (ObsID 300036010) will be publicly available in the archives after the proprietary period ends. The \nustar\ dataset  {(ObsID 30901010002)}
is already publicly available. 

\item[Code availability] The {\tt pion} photo-ionisation code is publicly available as part of the {\tt spex} package. The {\tt warmabs} photo-ionisation code is publicly available as part of the {\tt xstar} package. 
The {\tt ionabs} code is publicly available for download at
\url{https://github.com/ryotatomaru/Ionabs} 
as a local model for installation and use in the {\tt xspec} package. The {\sc xspec} model files used to make Extended Data Table 2 including the {\sc pion} tables are publically available for download at Zenodo\cite{Tomaru2025}
(\url{https://zenodo.org/records/15628497}).


\makeatletter
\apptocmd{\thebibliography}{\global\c@NAT@ctr 53\relax}{}{}
\makeatother

\setlength{\bibsep}{0.0pt}
{\sf\small

}

\item [Acknowledgments] 
This work was supported by JSPS KAKENHI grant numbers JP24KJ0152,JP22H00158, JP22H01268, JP22K03624, JP23H04899, JP21K13963, JP24K00638, JP24K17105, JP21K13958, JP21H01095, JP23K20850, JP24H00253, JP21K03615, JP24K00677, JP20K14491, JP23H00151, JP19K21884, JP20H01947, JP20KK0071, JP23K20239, JP24K00672, JP24K17104, JP24K17093, JP20K04009, JP21H04493, JP20H01946, JP23K13154, JP19K14762, JP20H05857, JP23H01211, JP23K03454, JP23K22548, JP23K03459, JP21H04493
and NASA grant numbers 80NSSC24K1148, 80NSSC24K1774, 80NSSC18K0978, 80NSSC20K0883, 80NSSC20K0737, 80NSSC24K0678, 80NSSC18K1684, 80NSSC25K7064, 80NSSC23K0995, 80NSSC18K0988, 80NSSC23K1656 and
80NSSC23K0684. 
CD acknowledges support from STFC through grant ST/T000244/1 and a Leverhulme Trust International Fellowship IF-2024-020. 
LC acknowledges support from NSF award 2205918. 
The material is based upon work supported by NASA under award number 80GSFC21M0002. 
This work was supported by the JSPS Core-to-Core Program, JPJSCCA20220002.
MM is supported by Yamada Science Foundation.
LG acknowledges financial support from Canadian Space Agency grant 18XARMSTMA. 
AT is supported in part by the Kagoshima University postdoctoral research program (KU-DREAM). 
SY acknowledges support by the RIKEN SPDR Program. 
IZ acknowledges partial support from the Alfred P. Sloan Foundation through the Sloan Research Fellowship. 
Part of this work was performed under the auspices of the U.S. Department of Energy by Lawrence Livermore National Laboratory under Contract DE-AC52-07NA27344. 
The material is based on work supported by the Strategic Research Center of Saitama University.

\item[Author contributions]
RT led the analysis, CD led the GX13+1 XRISM team. CD and JN were responsible for the manuscript writing. MM, FSP and SY led the XRISM Resolve data extraction, RT led the \nustar\ data, JN led the photoionisation code comparisons. The GX13+1 XRISM team (AK, EB, EC, MS, MDT, JN, HT, SY, LC, RS, KY, TD) contributed to discussions in regular team meetings. TK and
CK served as internal reviewers. The science goals of XRISM
were discussed and developed over 7 years by the XRISM Science Team, all
members of which are authors of this manuscript. All the instruments were
prepared by the joint efforts of the team. The manuscript was subject to an in-
ternal collaboration-wide review process. 

\item[Competing interests] The authors declare no competing interests

\item[Correspondence and requests for materials] should be addressed to 
Chris Done, Misaki Mizumoto, Joey Neilsen, or Ryota Tomaru \\ 

\item[Peer review information] {\it Nature} thanks Susmita Chakravorty, Knox Long, Daniel Proga and the 
other, anonymous, reviewer(s) for their contribution to the peer review of this work.

\item[ Reprints and permissions information] is available at \url{http://www.nature.com/reprints}.

\end{addendum}

\end{methods}


\large

\renewcommand{\arraystretch}{0.8}

\captionsetup[table]{justification=justified, singlelinecheck=off} 
\begin{table*}
\large
\baselineskip=24pt

\captionsetup{name=Extended Data Table}
{\caption{ \label{tab:ion_ion}Fit with ion-by-ion absorption plus scattered flux. }


\begin{tabular}{cccc} 
\hline 
continuum \\
\hline 
 {\tt diskbb} & $T_{\rm in} {\rm [keV]}$ & $ 1.56\pm0.01$ &\\
             & norm  
             & $66\pm2$&\\
{\tt bbody} & $kT {\rm [keV]}$ & $3.4^{+0.4}_{-0.3}$ &\\
            & norm $[\times 10^{-3}]$ & $4.5\pm 0.4$ &\\
{\tt gauss } &$ {\rm line~E~[keV]}$ & $6.4^{f}$& \\
 &$ \sigma {\rm ~[keV]}$ & $0.79\pm0.04$& \\
 &norm $[\times 10^{-2}]$ & $1.6^{+0.2}_{-0.1}  $ &\\

 \hline
 {\tt scatt} & $ f_{\rm scatt} $ &  $ 0.22 \pm0.01 $ &\\
 \hline 
 {\tt Ionabs} & $N_{\rm ion}~[{\rm 10^{18}~cm^{-2}}]$
 & $v_{\rm rms} ~[{\rm km~s^{-1}}]$ & $v_{\rm out}~[{\rm km~s^{-1}}]$\\ 
 \hline

 slow wind (16 ions)\\

 \hline
Si {\sc xxiv} & $0.6^{+0.8}_{-0.1}$ & $200.0^{+200.0}_{-160.0}$ & $-300.0^{+100.0}_{-80.0}$ \\
S {\sc xxvi} & $0.7^{+0.2}_{-0.1}$ & $110.0^{+20.0}_{-30.0}$ & $-370.0\pm20$ \\
Ar {\sc xviii} & $0.22^{+0.05}_{-0.02}$ & $120.0^{+40.0}_{-50.0}$ & $-340.0\pm20$ \\
Ca {\sc xix} & $0.021^{+0.005}_{-0.004}$ & $90.0^{+40.0}_{-60.0}$ & $-370.0\pm40$ \\
Ca {\sc xx} & $0.43^{+0.04}_{-0.03}$ & $120.0\pm 10.0$ & $-330.0\pm10.0$ \\
Ti {\sc xxii} & $0.04^{+0.01}_{-0.007}$ & $160.0^{+90.0}_{-100.0}$ & $-300.0\pm 100$ \\
Cr {\sc xxiii} & $0.12^{+0.02}_{-0.01}$ & $100.0\pm 20.0$ & $-320.0\pm 10.0$ \\
Cr {\sc xxiv} & $0.37^{+0.04}_{-0.03}$ & $120.0\pm20.0$ & $-320.0\pm 10.0$ \\
Mn {\sc xxiv} & $0.13\pm0.01$ & $160.0\pm20.0$ & $-330.0\pm20.0$ \\
Mn {\sc xxv} & $0.25\pm0.03$ & $160.0^{+20.0}_{-30.0}$ & $-330.0\pm20.0$ \\
Fe {\sc xxiv} & $0.34^{+0.05}_{-0.04}$ & $70.0\pm10.0$ & $-310.0\pm10.0$ \\
Fe {\sc xxv} & $20.0\pm2.0$ & $98.0\pm5.0$ & $-338.0^{+8.0}_{-7.0}$ \\
Fe {\sc xxvi} & $26.0^{+4.0}_{-3.0}$ & $117.0\pm5.0 $ & $-410.0\pm10.0$ \\
Co {\sc xxvi} & $0.09^{+0.02}_{-0.01}$ & $120.0^{+30.0}_{-40.0}$ & $-350.0\pm30.0$ \\
Ni {\sc xxvii} & $1.8^{+0.6}_{-0.3}$ & $80.0^{+20.0}_{-10.0}$ & $-300.0\pm20.0$ \\
Ni {\sc xxviii} & $2.0^{+3.0}_{-1.0}$ & $20.0^{+10.0}_{-20.0}$ & $-270.0\pm20.0$ \\
\hline
fast wind (4 ions)\\
\hline
Fe {\sc xxv} & $0.19^\pm0.03$ & $340.0^{+40.0}_{-50.0}$ & $-740.0^{+50.0}_{-60.0}$ \\
Fe {\sc xxvi} & $0.69^{+0.07}_{-0.06}$ & $440.0^{+50.0}_{-60.0}$ & $-1170.0\pm60.0$ \\
Ni {\sc xxvii} & $0.19\pm0.05$ & $260.0^{+40.0}_{-50.0}$ & $-590.0^{+70.0}_{-80.0}$ \\
Ni {\sc xxviii} & $0.5\pm0.08$ & $230.0^{+30.0}_{-40.0}$ & $-450.0^{+40.0}_{-50.0}$ \\
 \hline
  Fit$^b$ statistic & $\chi^2_\nu = 14189/13503$
  \vspace{12pt}

\end{tabular}

\begin{spacing}{1.0}
\footnotesize

$^a$ $N_{\rm ion,18}$ is the ion column density in units of $10^{18}$ cm$^{-2}$.

$^b$ Model is ${\tt TBabs\times (Ionabs^{16}_s Ionabs^{4}_f + f_{scatt} Ionabs^{4}_f)\times Int}$

$^f$ indicates fixed parameters. All other parameter 
errors are calculated at 90\% confidence levels.  

\end{spacing}

}
\end{table*}

\clearpage 

\large
\begin{table*}
\captionsetup{name=Extended Data Table}
{\caption{\label{tab:2pion_2emit_XR}  \textbf{Fit with two {\tt pion} absorbers plus their emission and scattered flux.}}} 

    \begin{tabular}{ccc}
    model name & parameter [unit] & value \\
    \hline
    &continuum&\\
    \hline
    {\tt diskbb} & $kT_{\rm in}~[{\rm keV}]$ & $1.52\pm 0.01$\\
    & norm  & $  430\pm40$\\
{\tt bbody} & $kT~[{\rm [keV]}]$ & $3.3\pm0.2$ \\
            & norm  & $2.6\pm0.3 \times 10^{-2}$ \\
{\tt gauss } &$ {\rm line~E~[keV]}$ & $6.4^{f}$ \\
 &$ \sigma {\rm ~[keV]}$ & $0.86\pm0.04$\\
 &norm & $1.0^{+0.8}_{-0.9}\times 10^{-1} $ \\
  \hline 

{\tt scatt}    &$f_{\rm scatt}$ &$3.7^{+0.6}_{-0.7}\times 10^{-2}$\\
 \hline
 & slow absorption and emission$^a$& \\
 \hline
 {\tt pion abs s}  &$\log (\xi/[{\rm erg~cm~s^{-1}}])$ &$3.88\pm0.01$\\
 & $N_{H} ~[10^{22} ~{\rm cm^{-2}}]$&$132^{+7}_{-8}$\\
&$v_{\rm rms}~[{\rm km~s^{-1}}]$ &$99\pm 3$\\
& $v_{\rm out}~[{\rm km~s^{-1}}]$& $-330\pm4$\\
 {\tt pion~emit s} & $v_{\rm rms}~[{\rm km~s^{-1}}]$& $600^{+200}_{-100}$ \\
  & norm & $5.2\pm0.9 \times 10^{-10}$ \\
\hline
& fast absorption and  emission$^a$&\\
\hline 
{\tt pion abs f} &$\log (\xi/[{\rm erg~cm~s^{-1}}])$ &$4.69^{+0.03}_{-0.04}$\\
 & $N_{H} ~[10^{22} ~{\rm cm^{-2}}]$&$79\pm9$\\
&$v_{\rm rms}~[{\rm km~s^{-1}}]$ &$400 \pm 20$\\
& $v_{\rm out}~[{\rm km~s^{-1}}]$& $-640\pm30$\\
 {\tt pion~emit f} & $v_{\rm rms}~[{\rm km~s^{-1}}]$& $750^{+60}_{-40} $ \\
  & norm & $1.4\pm0.2 \times 10^{-8}$ \\
\hline
Fit$^b$ statistic &  $\chi^2/\nu=  14339/13551$ &\\

\end{tabular}

\begin{flushleft}
\begin{spacing}{1.0}
\footnotesize
$^a$ The column density and ionisation state are tied between the absorption and emission. 

$^b$ model is ${\tt TBabs\times (abs_{s}abs_{f} Int +  
f_{\rm scatt} abs_{\rm f} Int + abs_{\rm s} emm_{\rm f} + f_{\rm scatt} emm_f + emm_{\rm s} )}$

$^f$ indicates fixed parameters. All other parameter 
errors are calculated at 90\% confidence levels. 

\end{spacing}
\end{flushleft}
\end{table*}

\end{document}